\documentclass[preprint,10pt,aps,showpacs]{revtex4}
\usepackage{amsmath,amsthm,amssymb}
\begin{document}
\title{An elliptic current operator for the eight-vertex model}
\author{Klaus Fabricius
\footnote{e-mail Fabricius@theorie.physik.uni-wuppertal.de}}
\affiliation
{ Physics Department, University of Wuppertal,
42097 Wuppertal, Germany}
\author{Barry~M.~McCoy
\footnote{e-mail mccoy@insti.physics.sunysb.edu}}
\affiliation{Institute for Theoretical Physics, State University of New York,
 Stony Brook,  NY 11794-3840}
\date{\today}
\preprint{YITPSB-06-19} 

\begin{abstract}

We compute the operator which creates the missing degenerate states in
the algebraic Bethe ansatz of the eight-vertex model at roots of unity and 
relate it to the concept of an elliptic current operator.
We find that in sharp contrast with the corresponding formalism in the 
six-vertex model at roots of unity \cite{fm4} the current operator is
not nilpotent with the consequence that in the construction of degenerate
eigenstates of the transfer matrix an arbitrary number of exact strings
can be added to the set of regular Bethe roots. Thus the original set of free
parameters $\{s,t\}$ of an eigenvector of the transfer matrix $T$ is enlarged to  
become $\{s,t,\lambda_{c,1},\cdots, \lambda_{c,n}\}$ with arbitrary string centers
$\lambda_{c,j}$ and arbitrary $n$.
\end{abstract}
\maketitle
\noindent
\section{Introduction}
\noindent
The spectrum of eigenvalues of the transfer matrix of the six-vertex
model and of the Hamiltonian of the $XXZ$ spin chain 
\begin{equation}
H_{XXZ}=-\sum_{j=1}^N\{\sigma_j^x\sigma_{j+1}^x+ 
\sigma_j^y\sigma_{j+1}^y +\cos \pi \gamma\sigma_j^z\sigma_{j+1}^z \}
\end{equation}
with periodic boundary conditions has many degenerate multiplets when
$\gamma$ is rational. Degenerate eigenvalues are
always a signal of extra symmetries of the system and the degeneracies
of the six-vertex model for rational $\gamma$ have been extensively 
analyzed \cite{dfm}-\cite{deg1} in terms of an $sl_2$ loop algebra symmetry.

The eight-vertex model of Baxter \cite{bax72}-\cite{Baxter2002} is 
a lattice model 
whose transfer matrix is given by
\begin{equation}
{\bf T}_8(v)|_{\mu,\nu}={\rm Tr}W(\mu_1,\nu_1)
\label{transfer}
W(\mu_2,\nu_2)\cdots W(\mu_N,\nu_N)
\end{equation}
where $\mu_j,\nu_j=\pm1$ and $W(\mu,\nu)$ is a $2\times 2$ matrix whose
non vanishing elements are given as
\begin{eqnarray}
W(+1,+1)|_{+1,+1}&=W(-1,-1)|_{-1,-1}
=\rho\Theta(2\eta)\Theta(\lambda-\eta)H(\lambda+\eta)=a(\lambda)
\nonumber\\
W(-1,-1)|_{+1,+1}&=W(+1,+1)|_{-1,-1}
=\rho\Theta(2\eta)H(\lambda-\eta)\Theta(\lambda+\eta)=b(\lambda)
\nonumber\\
W(-1,+1)|_{+1,-1}&=W(+1,-1)|_{-1,+1}
=\rho H(2\eta)\Theta(\lambda-\eta)\Theta(\lambda+\eta)=c(\lambda)
\nonumber\\
W(+1,-1)|_{+1,-1}&=W(-1,+1)|_{-1,+1}
=\rho H(2\eta)H(\lambda-\eta)H(\lambda+\eta)=d(\lambda)
\label{bw8}
\end{eqnarray}
where $H(u)$ and $\Theta(u)$ are Jacobi's Theta functions 
defined in  appendix A. 
In the following we set $\rho =1$.
This transfer matrix has a set of degeneracies similar to the 6
vertex model when $\eta$ is restricted to values $\eta_0$ given by 
\begin{equation}
\eta_0=2m_1 K/L
\label{root}
\end{equation}
where $L$ and $m_1$ are integers and
$K$ is the complete elliptic integral of the first kind.
Some of these degeneracies were recognized in ref.\cite{bax731} but
the full set seems to have only recently been obtained \cite{fm4}-\cite{fm5}. 
Corresponding
to the six-vertex model these degeneracies also indicate the existence
of a symmetry algebra over and beyond the star triangle equation 
\cite{bax72},\cite{baxb} which
guarantees the commutation of the transfer matrix for different values
of the spectral variable $\lambda$. This extra symmetry at rational values of
$\eta/K$ was first seen and exploited by Onsager \cite{ons} in his 1944
solution of the Ising model.

For the six-vertex model we have Chevalley generators \cite{dfm}, current
operators \cite{Odyssey}, and the degenerate states can be characterized by a
Drinfeld polynomial \cite{Odyssey},\cite{deg1}. 
For the eight-vertex model we know much
less. Deguchi \cite{deg2}-\cite{deg3} has produced  a
meromorphic function which should play the role of a Drinfeld polynomial
but because expressions for neither the Chevalley nor the current
generators are known it is not yet possible to make a connection of
this meromorphic function with a symmetry algebra.

The purpose of this paper is to partially fill this gap by finding the
operator which is the generalization of the current operator found for
the six-vertex model \cite{Odyssey} 
which we accordingly will call 'elliptic
current operator'. We will do this in analogy with our
previous computation \cite{Odyssey} for the six-vertex model by 
using the algebraic
Bethe ansatz of Takhtadzhan and  Faddeev \cite{TakFadd79} 
for the eight-vertex model.
This method is better adapted to the computation of the current operator than
the related formalism of Felder and Varchenko \cite{felder1},
\cite{felder2} used by Deguchi \cite{deg2}-\cite{deg3}.

In the description of degenerate eigenstates of the $T$ matrix the
concept of complete strings plays a central role. However there are two
quite different meanings of this term and 
to avoid confusion we will summarize here several of the significant
differences between the two concepts, the details of which will be
fully elaborated in the text below.
In a previous paper \cite{fm4} we explained the dimensions of 
degenerate subspaces
of $T$ by using the   $Q$-matrix constructed by Baxter in \cite{bax72}. 
This Q-matrix is defined
only when $\eta$ satisfies the root of unity condition 
$2L\eta=2m_1K+im_2K^{'}$. To distinguish this $Q$ matrix
from Baxter's  Q-matrix of ref. \cite{bax731,baxb} 
which exists for generic $\eta$ we denote it 
by $Q_{72}$. The eigenvalues of $Q_{72}$ are quasiperiodic with $2iK'$
as the imaginary quasiperiod. 
The eigenvalues of those eigenvectors of $Q_{72}$ 
which are degenerate eigenstates
of $T$ have string like subsets of zeros, which we will call $Q$-strings
\footnote{This definition rests on the properties of $Q_{72}$ shown in equ. (\ref{form2}).}.
These $Q$-strings have fixed string centers which have the property that 
for every string center $\lambda_c$ there is another
string center at $\lambda_c+iK'$. For each eigenstate the number of $Q$
strings is fixed \cite{fm4}.
For more details see section III, especially (\ref{form2}).

String like sequences of spectral parameters (\ref{lstring}) will
also occur as arguments in products of $B$-operators 
building up an eigenvector of $T$.  These eigenvectors are
quasiperiodic with an imaginary quasiperiod $iK'$.
These strings will be called $B$-strings. They were first reported
in \cite{bax731} and are further  discussed in \cite{Baxter2002}.
 The centers of $B$-strings
are free but because of the quasiperiodicity their imaginary parts are
restricted to lie between zero and $iK'$.
 Furthermore  we will show
below in sec. 5 that, in contrast with $Q$-strings, 
the number of $B$-strings used to produce  an eigenstate is not
fixed. These significant differences  between the two types of strings
mean that the two concepts are different and $Q$-strings cannot 
be considered as limiting or special cases of $B$ strings.

In the following we compute the creation operator for  $B$ strings. 
We give this result in 
sec. 2 as (\ref{BL1})-(\ref{Y}).
In sec. 3 we present results about the size of degenerate multiplets
in the eight-vertex model at roots of unity.
In sec. 4 we derive (\ref{BL1})-(\ref{Y}) and in sec. 5 we construct the
operator for multiple $B$ strings from the single $B$ string operator
(\ref{BL1}). We conclude in sec. 6 with a discussion of our results.

\section{Results} 
\noindent
In order to present our result for the elliptic current operator we
recall in appendix B the formalism and notations of the algebraic Bethe
ansatz used in  ref. \cite{TakFadd79} to compute some eigenvectors in the
root of unity case where (\ref{root}) holds. Here we explain why there
are states which are not given by this formalism and present our
result for the operator which creates these missing states. The details
of the computation of this creation operator are given in sec. 4.

Eigenvectors of the transfer matrix (\ref{transfer}) are given by (\ref{help100})
in Appendix B.
They describe all non degenerate eigenstates of $T$ and because of their dependence 
on parameters $s,t$ also some (but not all) degenerate eigenstates.
There are, however, eigenvectors where the root of unity condition 
(\ref{root}) holds for which the construction (\ref{help100}) is not
adequate. This happens if we consider a set of
$\lambda_k$ given by
\begin{equation}
\lambda_k=\lambda_c-(k-1)2\eta_0,\hspace{0.5 in}k=1, \cdots ,L_s
\label{lstring}
\end{equation}
where $\eta_0$ is given by (\ref{root}) and $\lambda_c$ is arbitrary.
Such a set of $\lambda_k$ is called a string and $\lambda_c$ is
called the string center. The string length $L_s$ is determined by the integer $L$
occurring in equ. (\ref{root}).
It will be shown that $L_s=L/2$ if $L$ is even and $L_s=L$ if $L$ is odd.
The attempt to construct eigenstates of $T$ containing 
$B$-strings by using (\ref{lstring}) in (\ref{help100}) fails because the
operator
\begin{equation}
B^{L_s}_{l}(\lambda_c)=B_{l+1,l-1}(\lambda_1)\cdots B_{l+L_s,l-L_s}(\lambda_{L_s})
\label{BL}
\end{equation}
is found numerically (in agreement with the analogous analytic computation of
Tarasov \cite{Tar} for the six-vertex model) to vanish
\begin{equation}
B^{L_s}_{l}(\lambda_c)=0
\label{stringz}
\end{equation}
for all $l,\lambda_c,s,t$.

Examples of states with a set of $\lambda_k$ given by (\ref{lstring})
have been known since the original work of
\cite{bax731}-\cite{bax733},\cite{fm4}-\cite{fm5} 
and thus
there must be a creation operator for these $B$-string states.  
This operator should be an elliptic generalization
of the loop algebra current found for the six-vertex model in \cite{Odyssey}.
In the following sections we shall prove that the creation operator 
of complete $B$-strings is
\begin{equation}
B^{L_s,1}_{l}(\lambda_c)=\sum_{j=1}^{L_s}B_{l+1,l-1}(\lambda_1)\cdots
\left({\frac{\partial B_{l+j,l-j}}{ \partial \eta}}(\lambda_{j})
-\hat{Z}_{j}{\frac{\partial B_{l+j,l-j}} {\partial \lambda}}(\lambda_{j})\right)
\cdots B_{l+L_s,l-L_s}(\lambda_{L_s}) 
\label{BL1}
\end{equation}
where
\begin{equation}
{\hat Z}_1(\lambda_c)={{\hat X}(\lambda_c)\over {\hat Y}(\lambda_c)}
\label{X}
\end{equation}
with
\begin{eqnarray}
&&{\hat X}(\lambda_c)=-2\sum_{k=0}^{L_s-1}k
\frac{\omega^{-2(k+1)}\rho_{k+1}}{P_kP_{k+1}}\\
&&{\hat Y}(\lambda_c)=
\sum_{k=0}^{L_s-1}\frac{\omega^{-2(k+1)}\rho_{k+1}}{P_kP_{k+1}}
\label{Y}
\end{eqnarray}
and
\begin{equation}
\hat{Z}_{j}(\lambda_c)=\hat{Z}_1(\lambda_c-(j-1)2\eta_0)
\end{equation}
where 
$\omega=e^{2\pi i m/L}$ is a Lth root of unity 
\begin{equation}
\rho_k=h^{N}(\lambda_c-(2k-1)\eta_0)\label{rho}
\end{equation}
and
\begin{equation}
P_k = \prod_{m=1}^{n_r}h(\lambda_c-\lambda^{r}_{m}-2k\eta_0).
\label{P}
\end{equation}
In (\ref{P}) the $\lambda^r_m$ (which we call regular Bethe roots) satisfy the
Bethe equation (\ref{Betheq}).

The functions ${\hat X}(\lambda_c)$ and ${\hat Y}(\lambda_c)$
generalize to the eight-vertex model the functions $X(\lambda_c)$ and
$Y(\lambda_c)$ previously obtained \cite{Odyssey} for the six-vertex model.
The function ${\hat Y}(\lambda_c)$ has been previously obtained by
Deguchi \cite{deg2}-\cite{deg3}.

\section{Rank of degenerate subspaces}
\noindent
We first present new exact results \cite{fm4} about the size of degenerate multiplets of
eigenvalues of the transfer matrix of the eight-vertex model.
It is known that in the six-vertex model at roots of unity the degenerate
subspaces of eigenstates of the transfer matrix are related to evaluation
representations of the $sl_2$ loop algebra. If the zeros of the Drinfeld polynomial
of a highest weight state have multiplicity 1 the dimension of
the multiplet is a power of 2. This is the generic case.
It is remarkable that the dimension of degenerate subspaces of the set of eigenstates
of the transfer matrix of the eight-vertex model for rational $\eta/K$ is also 
a power of 2. This has been shown in \cite{fm4} with quite a different method.
Because of its importance in the present context we give a short description
of the result in a more complete form.\\
In \cite{bax72} Baxter constructed a $Q$-matrix for the eight-vertex model under the 
restriction that
\begin{equation}
 2L\eta=2m_1K +im_2K'
\label{eta72}
\end{equation}
We  denote this $Q$-matrix here by $Q_{72}$ to distinguish it from the $Q$-matrix which Baxter 
defined in \cite{bax731} and  \cite{baxb}.
We note that there occurs a slightly different restriction on $\eta$ in the construction
of eigenstates of the eight-vertex transfer matrix \cite{bax731}-\cite{bax733,TakFadd79}
\begin{equation}
 L\eta=2m_1K +im_2K'
\label{eta73}
\end{equation}
In this work we consider only the case $m_2=0$.\\
In (\ref{eta72}) and (\ref{eta73}) $L,m_1,m_2$ are integers.We will have to use both restrictions 
depending on the context.
We have shown in  \cite{fm4} that $Q_{72}$ {\it does not exist if in (15) $L$ is odd and $m_1$ is even}.
$Q_{72}$ has the properties \cite{fm4}
\begin{eqnarray}
Q_{72}(v+2K)&=&(-1)^{\nu '}Q_{72}(v)\label{qper}\\
Q_{72}(v+2iK')&=&q^{-N}{\rm exp}(-iN\pi v/K)Q_{72}(v)
\label{qtrans}
\end{eqnarray}
where $\nu^{'}$ is the eigenvalue of the operator
\begin{equation}
S = \sigma_3\otimes \cdots \otimes \sigma_3
\end{equation}
which commutes with the transfer matrix $T$.
From this we derived the most general form for the eigenvalues of $Q_{72}$ \cite{fm4}:
\begin{equation}
Q_{72}(v)={\cal K}(q;v_k){\rm exp}(-i\nu\pi v/2K)\prod_{j=1}^{N}H(v-v_j)
\label{form1}
\end{equation}
The integers $\nu$ and $\nu^{'}$ satisfy 
\begin{equation}
\nu+\nu^{'}+N= {\rm even~~integer}
\label{sr1}
\end{equation}
\begin{equation}
N+(-\nu i K^{'} + \sum_{j=1}^{N}v_j)/K = {\rm even~~integer}
\label{sr2}
\end{equation}
which follows from equs. (\ref{qper}) and (\ref{qtrans}).
If subsets of roots form 
exact strings
 , the explicit form of (\ref{form1}) is
\begin{eqnarray}
Q_{72}(v)=\hat{{\cal K}}(q;v_k){\rm exp}(i(n_B-\nu)\pi v/2K)\prod_{j=1}^{n_B}
h(v-v^B_j)\nonumber\\
\times\prod_{j=1}^{n_L}H(v-iw_j)H(v-iw_j-2K/L)\cdots H(v-iw_j-2(L-1)K/L)
\label{form2}
\end{eqnarray}
\begin{equation}
2n_B+Ln_L=N.
\label{norootsa}
\end{equation}
$n_B$ is the number of Bethe roots $v_k$ and $n_L$ the number of exact $Q$-strings of length $L$.
{\it Note that $n_L$ is always even.}  
The eigenvalues of the transfer matrix are (equ.(C38) of \cite{bax72}):
\begin{equation}
t(v)Q_{72}(v) = h^{N}(v-\eta)Q_{72}(v+2\eta)+h^{N}(v+\eta)Q_{72}(v-2\eta)
\label{teig}
\end{equation}
or
\begin{equation}
t(v) = \exp(\frac{i\pi(n_B-\nu)\eta}{K})h^{N}(v-\eta)\prod_{j=1}^{n_B}\frac{h(v-v_j^{B}+2\eta)}{h(v-v_j^{B})}
+\exp(-\frac{i\pi(n_B-\nu)\eta}{K})h^{N}(v+\eta)\prod_{j=1}^{n_B}\frac{h(v-v_j^{B}-2\eta)}{h(v-v_j^{B})}
\label{teig1}
\end{equation}
where  according to equs. (\ref{sr2}) and (\ref{form2})
\begin{equation}
\nu =n_B+(2\sum_{j=1}^{n_B}Imv_j^{B}+L\sum_{j=1}^{n_L}w_j)/K^{'} 
\label{nu}
\end{equation}
It follows from a functional equation satisfied by the eigenvalues of $Q_{72}$ (see \cite{fm4} equs.(3.11)
and (3.12)) that for any given set of Bethe roots $v^{B}_k$ there are $2^{n_L}$ independent solutions $w_l$. 
Then
\begin{equation}
 w_j = w_j^{0}+\epsilon_j K^{'} \hspace{0.4 in}w_j^{0}< K^{'},\hspace{0.3 in}\epsilon_j=0,1, \hspace{0.2 in}j=1,\cdots n_L
\end{equation}
and 
\begin{equation}
\nu = n_B+\nu_0+L\sum_{j=1}^{n_L}\epsilon_j\hspace{0.2 in} {\rm with~~integer}~~ 
 \nu_0 =(2\sum_{j=1}^{n_B}Imv_j^{B}+L\sum_{j=1}^{n_L}w_j^{0})/K^{'}
\label{nu1}
\end{equation}
\begin{equation}
t(v) = \exp(-i\pi m\sum_{j=1}^{n_L}\epsilon_j)\hat{t}(v)
\label{hatt}
\end{equation}
\begin{equation}
\hat{t}(v) =
\exp(-i\pi \nu_0\eta/K)h^{N}(v-\eta)\prod_{j=1}^{n_B}\frac{h(v-v_j^{B}+2\eta)}{h(v-v_j^{B})}
+\exp(i\pi\nu_0)\eta/K)h^{N}(v+\eta)\prod_{j=1}^{n_B}\frac{h(v-v_j^{B}-2\eta)}{h(v-v_j^{B})}
\label{teig2}
\end{equation}
As $m$ is an odd integer we find that for a fixed set of Bethe roots $v_j^{B}, j=1 \cdots n_B$ 
the $2^{n_L}$ independent solutions $w_l$ give $2^{n_L-1}$ eigenstates of $T$ with eigenvalue $\hat{t}(v)$ and 
$2^{n_L-1}$ eigenstates of $T$ with eigenvalue $-\hat{t}(v)$.
$Q_{72}$ does not exist for even $m$ but we get a result also for even $m$ and odd $L$ using
\begin{equation}
t(v+K;K-\eta)=(-1)^{\nu^{'}}t(v;\eta)=\exp(-i\pi(\nu^{'}+ m\sum_{j=1}^{n_L}\epsilon_j)\hat{t}(v)
\label{tvK}
\end{equation}
It follows from equs. (\ref{sr1}) and (\ref{nu1}) that
\begin{equation}
n_B+\nu_0+L\sum_{j=1}^{n_L}\epsilon_j+\nu^{'} = 2r
\end{equation}
\begin{equation}
\nu^{'}+m\sum_{j=1}^{n_L}\epsilon_j = 2r-n_B-\nu_0+(m-L)\sum_{j=1}^{n_L}\epsilon_j
\label{sr3}
\end{equation}
As $m$ and $L$ are both odd the right hand side is either even or odd for {\it all} $2^{n_L}$ independent solutions $w_l$
Consequently the degenerate multiplet has $2^{n_L}$ elements.
This delivers much more detailed information on the size of degenerate multiplets than what was previously known.
The only precise information is found in \cite{bax731,bax732,bax733} where $2N$ special eigenvectors of $T$ are
constructed with degenerate subsets of size $2N/L$.
The main purpose of this paper is to describe a method to construct the complete set of eigenstates of
the transfer matrix corresponding to these $2^{n_L}$ sets of exact $Q$-strings.
It is well known that the eigenvectors of the transfer matrix of the eight-vertex model depend on parameters
$s,t$ as shown in \cite{bax733} and \cite{TakFadd79}. It shall be demonstrated in the following that this freedom
allows the construction of a small subset of degenerate states but is insufficient to generate the full degenerate
subspaces. 
\section{The root of unity limit}
\subsection{Exact $B$-strings in the eight-vertex model}
\noindent
We have shown that the length of complete $Q$- strings in the set of zeros of eigenvalues of $Q_{72}$ at $\eta=mK/L$ is $L$.
The results of Baxter \cite{bax731,bax732,bax733} and Takhtadzhan and Faddeev \cite{TakFadd79}
for the eight-vertex model are obtained for $\eta=2m_1K/L$ where $m_1,L$ are integers which include also
those $\eta$ for which $Q_{72}$ does not exist. 
We now determine the length of complete $B$-strings in this formalism.
Baxter was the first to discuss the existence of complete $B$-strings 
in the Bethe ansatz solution of the eight-vertex model at roots of unity 
$L\eta=2m_1 K$ (page 54 in \cite{bax733}). 
He defines them as strings of Bethe roots $u_j = u_c+2j\eta$ which will cancel out of the $TQ$-equation and Bethe's equations.
We apply this idea here to equ. (\ref{Betheq}).
The contribution $P_s$ of a string of length $L_s$
\begin{equation}
\lambda_k = \lambda_c-(k-1)2\eta_0, \hspace{0.4 in} k=1,\cdots ,L_s
\end{equation}
to the right hand side of
\begin{equation}
\frac{h^{N}(\lambda_j+\eta)}{h^{N}(\lambda_j-\eta)}=\exp(-4\pi im/L)
\frac{\prod_{k=1,k\neq j}^{n}h(\lambda_j-\lambda_k+2\eta_0)}
     {\prod_{k=1,k\neq j}^{n}h(\lambda_j-\lambda_k-2\eta_0)}
\label{BE}
\end{equation}
is after cancellations
\begin{equation}
P_{s}=
\frac{h(\lambda_j-\lambda_c+(L_s-1)2\eta_0)  h(\lambda_j-\lambda_c+L_s2\eta_0)}
{h(\lambda_j-\lambda_c-2\eta_0)h(\lambda_j-\lambda_c)}
\end{equation}
We find that
\begin{equation}
P_{s}= 1
\hspace{0.5 in} {\rm if}~~~ 2 L_s \eta_0=r 2K \hspace{0.5 in} 
\end{equation}
where $r$ is an integer.
\begin{equation}
2L_s m_1=rL
\end{equation}
As $(m_1,L)=1$ it follows $r=r_1m_1$ and $2L_s=r_1L$.\\
For {\it even L} we get the smallest possible $B$-string length $L_s=L/2$ by setting $r_1=1$ \\
Then $\eta=2m_1K/L=m_1K/L_s$ where $m_1$ is odd. We note that this is the case in which $Q_{72}$ exists. \\
For {\it odd L} we get the smallest possible $B$-string length $L_s=L$ by setting $r_1=2$.
Then $\eta=2m_1K/L_s$ and $Q_{72}$ does not exist.\\
The resulting string length is
\begin{equation}
L_s = L/2~~~ {\rm if}~~~ L={\rm even}, \hspace{0.3 in} L_s = L~~~ {\rm if}~~~ L={\rm odd}
\label{stringlength}
\end{equation}
In  the rest of this subsection we show that exact $B$-strings change the eigenvalue of the 
transfer matrix by a factor $\pm 1$ in agreement with the results (\ref{hatt})-(\ref{sr3}).
The eigenvalue of the transfer matrix is given in terms of Bethe roots 
\begin{equation}
\tilde{\Lambda} = \exp(2\pi im/L)h^{N}(\lambda+\eta)
\frac{\prod_{j=1,}^{n}h(\lambda-\lambda_j-2\eta_0)}
     {\prod_{j=1}^{n}h(\lambda-\lambda_j)}
+ \exp(-2\pi im/L)h^{N}(\lambda-\eta)
\frac{\prod_{j=1}^{n}h(\lambda-\lambda_j+2\eta_0)}
     {\prod_{j=1}^{n}h(\lambda-\lambda_j)}
\label{teig3}
\end{equation}
A $B$-string of length $L_s$ contributes to the first product in equ.(\ref{teig})
\begin{equation}
P_1 = \frac{h(\lambda-\lambda_c-2\eta_0)}{h(\lambda-\lambda_c-(L_s-1)2\eta_0)}
\end{equation}
and to the second product
\begin{equation}
P_2 = \frac{h(\lambda-\lambda_c-L_s2\eta_0)}{h(\lambda-\lambda_c)}
\end{equation}
We find that \\
$P_1=1$ and $P_2=1$ if $L_s2\eta_0=4K\times$ integer\\ 
and $P_1=-1$ and $P_2=-1$ if $L_s2\eta_0=2K\times$ odd integer.\\
The result is:\\
a)A single $B$-string changes the sign of the eigenvalue of $T$ if $L=$even.\\
b) For $L=$odd the sign of the eigenvalue is unchanged.\\
This result is  consistent with corresponding results obtained from the properties of $Q_{72}$ 
in the preceding subsection.
We recall that the number $Q$-strings in the set of eigenvalues of $Q_{72}$ is always even. In that case
the sign of the eigenvalue of $t(v)$ is determined by the factor ${\exp}(i(n_B-\nu)\pi v/2K)$ in (\ref{form2}).
$Q_{72}$ exists for $\eta={\rm (odd~~integer)/L_s}$ or for $m_1=2\times {\rm (odd~~integer)}, L=2L_s$.
In that case both signs of $t(v)$ occur. This corresponds to case a).
When $\eta={\rm (even~~integer)/(odd~~L_s)}$ we get from (\ref{tvK}) that there is no sign change like in
case b). \\
The results are summarized in table 1.\\ \\
\vspace{.2in}
Table 1. The string size and properties of $Q_{72}$ for $L\eta = 2m_1 K$ , $(m_1,L) = 1$.
The number of $L$-strings $n_L$ is defined in (\ref{norootsa}).
\vspace{.1in}

\begin{tabular}{|l|l|l|l|l|l|}\hline
$L$&$m_1$&$L_s$&$\eta$&$Q_{72}$&size of deg. subspace\\
\hline
odd&even or odd&$L$&even int. $\times$K/odd int.&does not exist&$2^{n_L}$\\
\hline
even&odd&$L/2$&odd int. $\times$K/integer&exists&$2^{n_L-1}$\\
\hline
\end{tabular}
\subsection{Construction of $B$-string creation operators.}
\noindent
We construct the operator $B_l^{L,1}$ for $\eta=\eta_0=2m_1K/L$ by 
using the formalism of the algebraic Bethe ansatz \cite{TakFadd79} by writing 
\begin{equation}
\eta=\eta_0+\epsilon
\end{equation}
and letting $\epsilon\rightarrow 0$.
Thus we consider the operator
\begin{equation}
\chi_{l}
=  B^{L}_{l}(\lambda_c)\prod_{m=L_s+1}^{n}B_{l+m,l-m}(\lambda_m)
\label{opchi}
\end{equation}
and the vector
\begin{equation}
\psi_{l}
=  \chi_{l}\Omega_{N}^{l-n}
\label{eigvec3}
\end{equation}
where $\lambda_j,~~j=L_s+1,\cdots n$ are at this stage 
arbitrary and where 
\begin{equation}
B^{L}_{l}(\lambda_c)=B_{l+1,l-1}(\lambda_1)\cdots B_{l+L_{s},l-L_{s}}(\lambda_{L_{s}})
\label{BL2}
\end{equation}
with arguments
\begin{equation}
\lambda_k
= \lambda_c-2(k-1)\eta_{0}-\hat{Z}_k(\lambda_c)\epsilon,~~~k=1,\cdots L_{s} 
\label{r}
\end{equation}
will give the desired creation operator when $\epsilon\rightarrow 0$.
The $B$-string length is given in equ. (\ref{stringlength}).
The allowed number of $B$ operators in (\ref{opchi}) will be determined later.
The key ingredient of our method is the function $\hat{Z}_{k}(\lambda_c)$ 
in equ.(\ref{r}). We do not obtain a result by
simply setting $\lambda_k= \lambda_c-2(k-1)(\eta_0+\epsilon)$.
The function $\hat{Z}_{k}(\lambda_c)$ has to be chosen such that 
$\sum_{l}\omega^{l} \psi_{l}$ (with appropriately defined sum) 
becomes an eigenvector of the transfer matrix.
The fundamental relations which allow the construction 
of eigenstates of the transfer matrix of the eight-vertex
model are derived using the commutations relations (\ref{C1})-(\ref{alpha})
\begin{eqnarray}
&&A_{l,l}(\lambda)\chi_{l}(\lambda_1,\cdots \lambda_n)
= \kappa(\lambda,\lambda_1,\cdots \lambda_n)
 \chi_{l-1}(\lambda_1,\cdots \lambda_n)A_{l+n,l-n}(\lambda)+\nonumber\\
&&\sum_{j=1}^{n}\kappa^{l}_{j}(\lambda,\lambda_1,\cdots \lambda_n)
\chi_{l-1}(\lambda_1,\cdots ,\lambda_{j-1}, 
\lambda, \lambda_{j+1}, \cdots, \lambda_n)A_{l+n,l-n}(\lambda_j)
\label{Avec}
\end{eqnarray}
\begin{eqnarray}
&&D_{l,l}(\lambda)\chi_{l}(\lambda_1,\cdots \lambda_n)
= \tilde{\kappa}(\lambda,\lambda_1,\cdots \lambda_n)
 \chi_{l+1}(\lambda_1,\cdots \lambda_n)D_{l+n,l-n}(\lambda)+\nonumber \\
&&\sum_{j=1}^{n}\tilde{\kappa}^{l}_{j}(\lambda,\lambda_1,\cdots \lambda_n)
\chi_{l+1}(\lambda_1,\cdots ,\lambda_{j-1},\lambda, \lambda_{j+1}, \cdots, \lambda_n)
 D_{l+n,l-n}(\lambda_j)
\label{Dvec}
\end{eqnarray}
The coefficients $\kappa,\tilde{\kappa},\kappa^{l}_{k},
\tilde{\kappa}^{l}_{k}$ are
\begin{equation}
\kappa(\lambda,\lambda_1,\cdots ,\lambda_n)
=\prod_{k=1}^{n}\alpha(\lambda,\lambda_k),
\hspace{0.25 in}
\tilde{\kappa}(\lambda,\lambda_1,\cdots ,\lambda_n)
=\prod_{k=1}^{n}\alpha(\lambda_k,\lambda)
\label{lam}
\end{equation}
\begin{equation}
\kappa^{l}_{j}(\lambda,\lambda_1,\cdots ,\lambda_n)=
-\beta_{l-1}(\lambda,\lambda_j)
 \prod_{k=1,k\neq j}^{n}\alpha(\lambda_j,\lambda_k)
\label{laml}
\end{equation}
\begin{equation}
\tilde{\kappa}^{l}_{j}(\lambda,\lambda_1,\cdots ,\lambda_n)=
 \beta_{l+1}(\lambda,\lambda_j)
\prod_{k=1,k\neq j}^{n}\alpha(\lambda_k,\lambda_j)
\label{tlaml}
\end{equation}
We must carefully expand (\ref{eigvec3}) and (\ref{lam})-(\ref{tlaml})
to first order as $\epsilon\rightarrow 0$.\\
The expansion of $B_{k,l}(\lambda)$ is
\begin{equation}
B_{l+m,l-m}(s,t,\lambda_m,\eta)=B_{l+m,l-m}(s,t,\lambda_{c},\eta_0)
+\epsilon \left({\partial B_{l+m,l-m}\over \partial \eta}
(s,t,\lambda_{c},\eta_0)
-\hat{Z}_m{\partial  B_{l+m,l-m}\over \partial \lambda}
(s,t,\lambda_{c},\eta_0)\right)
\label{bkle}
\end{equation}
if $\lambda_m$ is of type (\ref{r}).
\subsection{The expansion of $\kappa$ and $\kappa^{l}_{k}$}
\noindent
The coefficients 
$\kappa(\lambda,\lambda_1,\cdots ,\lambda_n)$
and
$\tilde{\kappa}(\lambda,\lambda_1,\cdots ,\lambda_n)$
are in the limit $\epsilon \rightarrow 0$ given by
\begin{equation}
\kappa(\lambda,\lambda_1,\cdots ,\lambda_n)
=(-1)^{L+1}\kappa^{r}(\lambda,\lambda_{L_{s}+1},\cdots ,\lambda_n)
\hspace{0.25 in}
\tilde{\kappa}(\lambda,\lambda_1,\cdots ,\lambda_n)
=(-1)^{L+1}\tilde{\kappa}^{r}(\lambda,\lambda_{L_{s}+1},\cdots ,\lambda_n)
\label{Lambda1}
\end{equation}
\begin{equation}
\kappa^{r}(\lambda,\lambda_{L_{s}+1},\cdots ,\lambda_n)
=\prod_{k=L_s+1}^{n}\alpha(\lambda,\lambda_k)
\hspace{0.25 in}
\tilde{\kappa^{r}}(\lambda,\lambda_{L_{s}+1},\cdots ,\lambda_n)
=\prod_{k=L_s+1}^{n}\alpha(\lambda_k,\lambda)
\label{Lambda2}
\end{equation}
as the $B$-string part becomes $\pm 1$ because of periodicity. 
This guarantees that $B^{L,1}_{l}$ if applied to an eigenstate
does change the eigenvalue at most by changing its sign.
The superscript $r$ indicates that $\kappa^{r}$ and $\tilde{\kappa^{r}}$
depend only on those arguments $\lambda_j,j=L_s+1, \cdots n$ which finally will be set
to be regular Bethe roots.\\
We further observe that on the right hand side of (\ref{laml})
all factors are of order $\epsilon^{0}$ 
except the factor $\alpha(\lambda_k,\lambda_{k+1})=O (\epsilon^{1})$ 
with $1\leq k\leq L_s$
and equivalently on the right hand side of (\ref{tlaml})
all factors are of order $\epsilon^{0}$ 
except $\alpha(\lambda_{k-1},\lambda_{k})=O(\epsilon^{1})$ with $1\leq
k \leq L_s$.
It follows for $1\leq  k \leq L_s$ that
\begin{equation}
\kappa^{l}_{k}(\lambda)=\epsilon(-1)^{L+1}(\hat{Z}_{k+1}-\hat{Z}_{k}-2)
\frac{h^{'}(0)h(\tau_{l-1}+\lambda_{0k}-\lambda)}
{h(\tau_{l-1})h(\lambda_{0k}-\lambda)}
\frac{\prod_{m=L_s+1}^{n}h(\lambda_{0k}-\lambda_{m}-2\eta_0)}
{\prod_{m=L_s+1}^{n}h(\lambda_{0k}-\lambda_{m})}
\label{llexpand}
\end{equation}
\begin{equation}
\tilde{\kappa}^{l}_{k}(\lambda)=-(-1)^{L+1}\epsilon(\hat{Z}_{k}-\hat{Z}_{k-1}-2)
\frac{h^{'}(0)h(\tau_{l+1}+\lambda_{0k}-\lambda)}
{h(\tau_{l+1})h(\lambda_{0k}-\lambda)}
\frac{\prod_{m=L_s+1}^{n}h(\lambda_{0k}-\lambda_{m}+2\eta_0)}
{\prod_{m=L_s+1}^{n}h(\lambda_{0k}-\lambda_{m})}
\label{tllexpand}
\end{equation}
The variables
\begin{equation}
\lambda_{0k}=\lambda_c-2(k-1)\eta_0,~~~k=1, \cdots ,L_s
\end{equation}
are the $B$-string arguments and the variables $\lambda_m, m=L_s+1, \cdots , n$ are still 
arbitrary and will be chosen finally as regular roots.
If however $k>L_s$  the factor related to the $B$-string is $(-1)^{L+1}$ and the limit
is of order $\epsilon^{0}$
\begin{equation}
\kappa^{l}_{k}(\lambda,\lambda_1,\cdots ,\lambda_n)=
(-1)^{L+1}\kappa^{r,l}_{k}(\lambda,\lambda_{L_{s}+1},\cdots ,\lambda_n)
\hspace{0.3 in}
\tilde{\kappa}^{l}_{k}(\lambda,\lambda_1,\cdots ,\lambda_n)=
(-1)^{L+1}\tilde{\kappa}^{r,l}_{k}(\lambda,\lambda_{L_{s}+1},\cdots,\lambda_n)
\end{equation}
\begin{equation}
\kappa^{r,l}_{k}(\lambda,\lambda_{L_{s}+1},\cdots ,\lambda_n)=
-\beta_{l-1}(\lambda,\lambda_k)
\prod_{j=L_s+1,j\neq k}^{n}\alpha(\lambda_k,\lambda_j)
\hspace{0.3 in}
\tilde{\kappa}^{r,l}_{k}(\lambda,\lambda_{L_{s}+1},\cdots ,\lambda_n)=
 \beta_{l+1}(\lambda,\lambda_k)
\prod_{j=L_s+1,j\neq k}^{n}\alpha(\lambda_j,\lambda_k)
\end{equation}

\subsection{Expansion of state vectors}
\noindent
We may now expand the vector $\psi_l$ by using the expansion (\ref{bkle}) of
$B_{k,l}$ in (\ref{eigvec3}) to find
that  $\psi_{l}=\psi_{l}^{(0)}+\epsilon \psi_{l}^{(1)}$ where
\begin{equation}
\psi_{l}^{(1)}=  B^{L,1}_{l}(\lambda_c)\prod_{m=L_s+1}^{n}
B_{l+m,l-m}(\lambda_m)\Omega_{N}^{l-n}
\label{psi1}
\end{equation}
and
\begin{equation}
B^{L,1}_{l}(\lambda_c)=\sum_{j=1}^{L_s}B_{l+1,l-1}(\lambda_1)\cdots
\left({\partial B_{l+j,l-j}\over \partial \eta}(\lambda_{j})
-\hat{Z}_j{\partial B_{l+j,l-j}\over \partial \lambda}(\lambda_{j})\right)
\cdots B_{l+L_s,l-L_s}(\lambda_{L_s}) 
\end{equation}
The allowed number of operators $B$ in (\ref{psi1}) follows from the
periodicity properties of $A_{k,l},\cdots , D_{k,l}$ for $\eta=\eta_0=2m_1K/L$.
Their period in $k,l$ is $L$. It follows that
\begin{equation}
A_{l+n,l-n}(\lambda)\Omega^{l-n}_N = A_{l+n+r_1L,l-n-r_2L}(\lambda)\Omega^{l-n}_N 
= h^N(\lambda+\eta)\Omega^{l-1-n}_N
\end{equation}
and
\begin{equation}
D_{l+n,l-n}(\lambda)\Omega^{l-n}_N = D_{l+n+r_1L,l-n-r_2L}(\lambda)\Omega^{l-n}_N 
= h^N(\lambda-\eta)\Omega^{l+1-n}_N
\end{equation}
if
\begin{equation}
2n+(r_1+r_2)L = N
\label{noroots}
\end{equation}
for integers $r_1,r_2$
When we insert (\ref{Lambda1}) and (\ref{llexpand}) into (\ref{Avec}) we find
\begin{eqnarray}
&&A_{l,l}(\lambda)\psi_{l}^{(1)}
=(-1)^{L+1}\Lambda(\lambda)\psi_{l-1}^{(1)}+(-1)^{L+1}\frac{h^{'}(0)}{h(\tau_{l-1})}
\sum_{k=1}^{L_s}(\hat{Z}_{k+1}-\hat{Z}_{k}-2)
 \frac{h(\tau_{l-1}+\lambda_{0k}-\lambda)}{h(\lambda_{0k}-\lambda)} 
\nonumber \\
&&\times h^{N}(\lambda_{0k}+\eta_0)
\frac{\prod_{m=L_s+1}^{n}h(\lambda_{0k}-\lambda_{m}-2\eta_0)}
{\prod_{m=L_s+1}^{n}h(\lambda_{0k}-\lambda_{m})}
\psi_{l-1}(\lambda_{01},\cdots, \lambda_{0,k-1},
\lambda, \lambda_{0,k+1},\cdots,\lambda_{0L_s},\cdots ,\lambda_{n})\nonumber \\
&&+\sum_{k=L_s+1}^{n}\Lambda^{l}_{k}\psi^{(1)}_{l-1}
(\lambda_{01},\cdots, \lambda_{0L_s},\cdots \lambda_{k-1},
\lambda, \lambda_{k+1},\cdots,\lambda_{n})
\label{AllPsi}
\end{eqnarray}
and when we insert (\ref{Lambda1}) and (\ref{tllexpand})  into (\ref{Dvec}) 
we obtain
\begin{eqnarray}
&&D_{l,l}(\lambda)\psi_{l}^{(1)}
=  (-1)^{L+1}\tilde{\Lambda}(\lambda)\psi_{l+1}^{(1)}
-(-1)^{L+1}\frac{h^{'}(0)}{h(\tau_{l+1})}
\sum_{k=1}^{L_s}(\hat{Z}_{k}-\hat{Z}_{k-1}-2)
\frac{h(\tau_{l+1}+\lambda_{0k}-\lambda)}{h(\lambda_{0k}-\lambda)} \nonumber \\
&&\times h^{N}(\lambda_{0k}-\eta_0)
\frac{\prod_{m=L_s+1}^{n}h(\lambda_{0k}-\lambda_{m}+2\eta_0)}
{\prod_{m=L_s+1}^{n}h(\lambda_{0k}-\lambda_{m})}
\psi_{l+1}(\lambda_{01},\cdots, \lambda_{0,k-1},
\lambda, \lambda_{0,k+1},\cdots,\lambda_{0L_s},\cdots ,\lambda_{n})\nonumber\\
&&+\sum_{k=L_s+1}^{n}\tilde{\Lambda}^{l}_{k}\psi^{(1)}_{l+1}(\lambda_{01},
\cdots, \lambda_{0L_s},\cdots, \lambda_{k-1},
\lambda, \lambda_{k+1},\cdots,\lambda_{n})
\label{DllPsi}
\end{eqnarray}
where 
\begin{equation}
\Lambda(\lambda)=
h^{N}(\lambda+\eta)\kappa^{r}(\lambda,\lambda_{L_s+1},\cdots ,\lambda_n)
\hspace{0.3 in}
\tilde{\Lambda}(\lambda)=
h^{N}(\lambda-\eta)\tilde{\kappa}^{r}(\lambda,\lambda_{L_s+1},\cdots ,\lambda_n)
\hspace{0.3 in}
\end{equation}
\begin{equation}
\Lambda^{l}_{k}(\lambda)=
(-1)^{L+1}h^{N}(\lambda+\eta)\kappa^{r,l}_{k}(\lambda,\lambda_{L_{s}+1},\cdots ,\lambda_n)
\end{equation}
\begin{equation}
\tilde{\Lambda}^{l}_{k}(\lambda)=
(-1)^{L+1}h^{N}(\lambda-\eta)\tilde{\kappa}^{r,l}_{k}(\lambda,\lambda_{L_{s}+1},\cdots ,\lambda_n).
\end{equation}
We remark that on the right hand sides of (\ref{AllPsi}) and (\ref{DllPsi}) in the first sum 
$\psi_{l-1}(\lambda_{01},\cdots, \lambda_{0,k-1},
\lambda, \lambda_{0,k+1},\cdots,\lambda_{n})$
is $O(\epsilon^{0})$ because there is not a complete $B$-string:
$\lambda_{0k}$ is replaced by $\lambda$.
In the last line however where $k>L_s$ the $B$-string is 
complete and therefore the first order term $\psi^{(1)}_{l-1}$
has to be taken.

We add (\ref{AllPsi}) and (\ref{DllPsi}),multiply by
$\omega^{l}=\exp(2\pi iml/L)$ and sum over $l$.
After the shift  $l \rightarrow l+1$  in the 
first sum and $l \rightarrow l-1$ in the second sum we get
\begin{eqnarray}
&&T\sum_{l=0}^{L-1}\omega^l\psi^{(1)}_l=
\sum_{l=0}^{L-1}(A_{l,l}(\lambda)
+D_{l,l}(\lambda))\omega^{l}\psi_{l}^{(1)}\nonumber\\
&& = (-1)^{L+1}(\omega\Lambda+\omega^{-1}\tilde{\Lambda})\sum_{l=0}^{L-1}\omega^{l}\psi_{l}^{(1)}+
(-1)^{L+1}h^{'}(0)\sum_{l=0}^{L-1}\omega^{l}
\sum_{k=1}^{L_s}\frac{h(\tau_{l}+\lambda_{0k}-\lambda)}
{h(\tau_{l})h(\lambda_{0k}-\lambda)}\nonumber \\
&&[\omega(\hat{Z}_{k+1}-\hat{Z}_{k}-2)
h^{N}(\lambda_{0k}+\eta_0)
\frac{\prod_{m=L_s+1}^{n}h(\lambda_{0k}-\lambda_m-2\eta_0)}
{\prod_{m=L_s+1}^{n}h(\lambda_{0k}-\lambda_m)}- \nonumber\\
&&\omega^{-1}(\hat{Z}_{k}-\hat{Z}_{k-1}-2)
h^{N}(\lambda_{0k}-\eta_0)
\frac{\prod_{m=L_s+1}^{n}h(\lambda_{0k}-\lambda_m+2\eta_0)}
{\prod_{m=L_s+1}^{n}h(\lambda_{0k}-\lambda_m)}]\nonumber\\
&&\psi_{l}(\lambda_{01},\cdots, \lambda_{0,k-1},\lambda, 
\lambda_{0,k+1},\cdots,\lambda_{L_s},\cdots ,\lambda_{n})\nonumber\\
&&+\sum_{l=0}^{L-1}\omega^{l}\sum_{k>L_s}(\omega\Lambda^{l+1}_{k}+\omega^{-1}\tilde{\Lambda}^{l-1}_{k})
\psi_{l}^{(1)}(\lambda_{01},\cdots,\lambda_{0L_s},\cdots , 
\lambda_{k-1},\lambda, \lambda_{k+1},\cdots,\lambda_{n})
\label{AD1}
\end{eqnarray}

\subsection{Determination of $\hat{Z}_{k}$}
\noindent
In order that $\sum_{l=0}^{L-1}\omega^l\psi_l$ be an eigenvector all
terms on the right hand side of (\ref{AD1}) must vanish except the first. The arguments of
\cite{TakFadd79} show that the last term will vanish
if the $\lambda_k, k=L_s+1,\cdots ,n$ are chosen to satisfy the Bethe's equation
(\ref{Betheq}).
The remaining unwanted terms in equ. (\ref{AD1}) will vanish if we set
\begin{eqnarray}
&&\omega(\hat{Z}_{k+1}-\hat{Z}_{k}-2)h^{N}(\lambda_{c}-(2k-3)\eta_0)
\prod_{m=L_s+1}^{n}h(\lambda_{c}-\lambda_{m}-2k\eta_0)=
 \nonumber \\
&&\omega^{-1}(\hat{Z}_{k}-\hat{Z}_{k-1}-2)h^{N}(\lambda_c-(2k-1)\eta_0)
\prod_{m=L_s+1}^{n}h(\lambda_c-\lambda_{m}-2(k-2)\eta_0) 
\label{XX}
\end{eqnarray}
To simplify the notation we recall the definitions of $\rho_k$
 and $P_k$ in (\ref{rho}) and (\ref{P}) and further define
\begin{equation}
f_k = \hat{Z}_{k+1}-\hat{Z}_{k}-2
\end{equation}
where 
\begin{equation}
\hat{Z}_k=\hat{Z}_1(\lambda-2(k-1)\eta_0)
\label{Xk}
\end{equation}
Thus (\ref{XX}) is rewritten as
\begin{equation}
\omega f_k\rho_{k-1}P_k = \omega^{-1}f_{k-1}\rho_kP_{k-2}
\end{equation}
We define $\hat{f}_k=\omega^{2k}f_k$ and get
\begin{equation}
\hat{f}_k\rho_{k-1}P_k = \hat{f}_{k-1}\rho_kP_{k-2}
\end{equation}
\begin{equation}
\frac{\hat{f}_k}{\hat{f}_{k-1}}= \frac{\rho_kP_{k-2}}{\rho_{k-1}P_k}=\frac{\rho_kP_{k-2}P_{k-1}}{\rho_{k-1}P_{k-1}P_k}
\end{equation}
It follows that
\begin{equation}
\hat{f}_k = \tilde{g}\frac{\rho_k}{P_{k-1}{P_k}}
\label{fk1}
\end{equation}
and
\begin{equation}
f_k = \tilde{g}\frac{\omega^{-2k}\rho_k}{P_{k-1}{P_k}}
\label{fk2}
\end{equation}
$\tilde{g}$ follows from $\sum_k \hat{f}_k= -2L_s$ :
\begin{equation}
\tilde{g} = -\frac{2L_s}{\sum_{k=0}^{L_s-1}\frac{\omega^{-2k}\rho_k}{P_{k-1}P_{k}}}=
 -\frac{2L_s}{\sum_{k=0}^{L_s-1}\frac{\omega^{-2(k+1)}\rho_{k+1}}{P_{k}P_{k+1}}}
\end{equation}
As $Q_k=\frac{\rho_k}{P_{k-1}P_{k}}$  satisfies $Q_{k+L_s}=Q_{k}$ we find that 
\begin{equation}
\tilde{g}(\lambda+2\eta_0) = \omega^{2}\tilde{g}(\lambda)
\end{equation}
It follows from equ.(\ref{fk2}) that
\begin{equation}
\hat{Z}_1-\hat{Z}_0-2 = \tilde{g}\frac{\rho_0}{P_{-1}P_0}
\label{X10}
\end{equation}
We make the ansatz
\begin{equation}
\hat{Z}_1(\lambda) = \tilde{g}\sum_{k=0}^{L_s-1}c_k\frac{\rho_{k+1}}{P_kP_{k+1}}
\end{equation}
It follows
\begin{equation}
\hat{Z}_0(\lambda) = \hat{Z}_1(\lambda+2\eta_0) = \tilde{g}(\lambda+2\eta_0)\sum_{k=0}^{L_s-1}c_{k+1}\frac{\rho_{k+1}}{P_kP_{k+1}}=
\omega^{2} \tilde{g}(\lambda)\sum_{k=0}^{L_s-1}c_{k+1}\frac{\rho_{k+1}}{P_kP_{k+1}}
\end{equation}
and
\begin{equation}
\hat{Z}_1-\hat{Z}_0-2 = \tilde{g}\sum_{k=0}^{L_s-1}(c_k-\omega^2c_{k+1}+\frac{1}{L_s}
\omega^{-2(k+1)})\frac{\rho_{k+1}}{P_{k}P_{k+1}}
\label{X10a}
\end{equation}
Comparison of (\ref{X10}) with (\ref{X10a}) gives 
\begin{equation}
c_k-\omega^2c_{k+1}+\frac{1}{L_s}\omega^{-2(k+1)}=0~~~,k=0,\cdots ,L_s-2
\label{c1}
\end{equation}
and
\begin{equation}
c_{L_s-1}-\omega^{2}c_{L_s}=1-\frac{1}{L_s}
\label{c2}
\end{equation}
We define $c_k=\omega^{-2(k+1)}\hat{c}_k$.
Equ. (\ref{c1}) and (\ref{c2}) are then simplified to
\begin{equation}
\hat{c}_k-\hat{c}_{k+1}=-\frac{1}{L_s}~~~,k=0,\cdots ,L_s-2
\label{c1a}
\end{equation}
and
\begin{equation}
\hat{c}_{L{_s}-1}-\hat{c}_{L_s}=1-\frac{1}{L_s}
\label{c2a}
\end{equation}
This gives the coefficients
\begin{equation}
c_k = \omega^{-2(k+1)}\frac{k}{L_s} \hspace{0.3 in}k=0,\cdots,L_s-1,~~~c_{L_s}=c_0
\label{co}
\end{equation}
\begin{equation}
\hat{Z}_1=-2\frac{\sum_{k=0}^{L_s-1}k\frac{\omega^{-2(k+1)}\rho_{k+1}}{P_kP_{k+1}}}
{\sum_{k=0}^{L_s-1}\frac{\omega^{-2(k+1)}\rho_{k+1}}{P_kP_{k+1}}}
\label{X1}
\end{equation}
Thus we have obtained the desired result (\ref{BL1}) for $B^{L,1}_{l}.$

\section{Multiple $B$-strings}
\noindent
We use a more compact notation to show the essentials of 
the proof that the result (\ref{X1}) and (\ref{BL1})
is correct also for multiple $B$-strings.
Let $\psi_2$ denote a state with two substrings like (\ref{BL}).
\begin{equation}
\psi_2 = B^{L}(\lambda_{c_1})B^{L}(\lambda_{c_2})B^{reg}
\end{equation}
We first show that equ.(\ref{Avec}) is then of 
order $\epsilon^{2}$. It is sufficient to show this for
\begin{equation}
\sum_{k=1}^{n}\Lambda^{l}_{k}(\lambda,\lambda_1,\cdots \lambda_n)
\psi_{l-1}(\lambda_1,\cdots ,\lambda_{k-1}, \lambda, \lambda_{k+1}, 
\cdots, \lambda_n)
\label{s2}
\end{equation}
We write 
\[
 \Lambda^{l}_{k}(\lambda,\lambda_1,\cdots ,\lambda_n)=
-\beta_{l-1}(\lambda,\lambda_k)h^{N}(\lambda+\eta)\prod_{j=1,j\neq k}^{n}\alpha(\lambda_k,\lambda_j)=p_1p_2p_{reg}
\]
\[
p_1=\prod_{j=1,j\neq k}^{L_s}\alpha(\lambda_k,\lambda_j) \hspace{0.3 in}
p_2=\prod_{j=L_s+1,j\neq k}^{2L_s}\alpha(\lambda_k,\lambda_j)\hspace{0.3 in}
p_{reg}=\prod_{j=2L_s+1,j\neq k}^{n}\alpha(\lambda_k,\lambda_j)
\]
Each term of the sum in (\ref{s2}) has the form
\[
\Lambda^{l}_{k}(\lambda,\lambda_1,\cdots \lambda_n)
\psi_{l-1}(\lambda_1,\cdots ,\lambda_{k-1}, \lambda, \lambda_{k+1}, \cdots, \lambda_n) \sim
p_1p_2p_{reg}B^{L}(\lambda_{c_1})B^{L}(\lambda_{c_2})B^{reg}
\]
In the limit $\epsilon \rightarrow 0$ the order of the power of $\epsilon$
depends on the position of $k$.
{
\begin{tabbing}
     xxxxxxx    \=          xxxxxxxx    \=          xxxxxxxx    \=     xxxxxxxx    \=xxxxxxx      \= xxxxxxx    \=        \kill
          \>$p_1$\>$p_2$\>$p_{reg}$\>$B^{L}(\lambda_{c_1}$)\>$B^{L}(\lambda_{c_2})$\>$B^{reg}$\\
 k $\in$ $S_1$ \> ($\epsilon^{1}$)\> ($\epsilon^{0}$)\> ($\epsilon^{0}$)\> ($\epsilon^{0}$)\> ($\epsilon^{1}$)\> ($\epsilon^{0}$)\\
 k $\in$ $S_2$ \> ($\epsilon^{0}$)\> ($\epsilon^{1}$)\> ($\epsilon^{0}$)\> ($\epsilon^{1}$)\> ($\epsilon^{0}$)\> ($\epsilon^{0}$)\\
 k $\in$ $R_0$ \> ($\epsilon^{0}$)\> ($\epsilon^{0}$)\> ($\epsilon^{0}$)\> ($\epsilon^{1}$)\> ($\epsilon^{1}$)\> ($\epsilon^{0}$)\\
\end{tabbing}
}
where $S_1 = \{1,2,\cdots ,L_s\}$,$S_2 = \{L_s+1,L_s+2,\cdots ,2L_s\}$, $R_0 = \{2L_s+1,2L_s+2,\cdots ,n\}$.\\
It is evident that equ. (\ref{Avec}) is then of second order. It also follows that terms of the first and
second lines of the preceding list are removed by (\ref{X1}) and terms from the last line
add up to zero if the variables $\lambda_k$
for  $k \in R_0$ are regular roots.\\
The total number of operators $B$ building an eigenvector is restricted by
\begin{equation}
2(n_B+n_s) + rL = N
\end{equation}
(see equ. (\ref{noroots}))
where $n_B$ is the number of regular roots and $n_s$ is the number of roots
belonging to $B$-strings. It is important to note that the 
integer $r$ may be positive
and negative. It follows that there is no restriction on the number of 
$B$-string-operators in a state vector. To elucidate the role of $B$-strings we note
that the analytical expression for eigenstates of the eight-vertex model
depends on two free parameters $s,t$. For degenerate eigenstates which form a 
space of dimension $d$ this means that a subspace of dimension $d_0 < d$
can be constructed by the variation of $s$ and $t$ without applying the elliptic 
current operator.
Detailed numerical studies have revealed that the variation of s,t will 
only give the full degenerate eigenspace for very small d (e.g. d=2).
In all other cases one needs to use
the elliptic current operator with the additional freedom to choose the string center 
to generate the full subspace. After this is achieved by adding a certain number of 
$B$-strings (the exact number depends on the system size $N$ and the value of $\eta$) 
the addition of more and more $B$-strings will only map this subspace into itself.
In particular adding $B$- strings to a singlet state with
$n_B=N/2$ Bethe-roots does not destroy this state but reproduces it.
It is not this redundancy which is important, but the prospect that this
might have to do with a cyclic nature of the hidden symmetry.
Finally we stress that the addition of $B$-strings does not single out a special basis
as the string centers can be chosen randomly.\\
The numerical tests have been performed for spin chains of length
$ N=6,8,10,12 $ and crossing parameters $\eta=K/2, K/3, 2K/3$.

\section{Discussion of the result}
\noindent
In this paper we have constructed the creation operator of $B$-strings in the
algebraic Bethe ansatz of the eight-vertex model. We have made no
connection with a symmetry algebra and therefore we are unable to even
define what is meant by a current operator.Nevertheless we have used
the phrase ``elliptic current operator'' in the title because the operator 
 $B_{l}^{L,1}(\lambda_c)$ is obviously a generalization of the current
operator $B^{(L)}_{6}(v)$ of the six vertex model at roots of unity\cite{Odyssey}.
The term 'generalization' is used here admittedly in a vague sense.
The true relation of the operator (\ref{BL1}) to the six-vertex current operator
derived in ref. \cite{Odyssey} can only be found by a careful analytical investigation
of the trigonometric limit of (\ref{BL1}). The six-vertex limit of eigenvectors
of the eight-vertex model with B-string number $0$ is presented in great detail
in ref.\cite{DdVG} which will presumably be of great help also in understanding 
the limit of the $B$-string operator (\ref{BL1}).\\
The operator  $B_{l}^{L,1}$ serves as generator of missing 
eigenstates in the eigenspaces of degenerate eigenvalues of $T$. 
The eigenstates (\ref{help100}) of the algebraic Bethe
ansatz depend on parameters $s,t$. However, in the case that an eigenvalue 
is degenerate one obtains only a subspace of 
the related degenerate eigenspace by varying $s,t$ in
(\ref{help100}). 
The complete set of missing states is found 
by adding a sufficient number of B-strings. This is similar to the string 
operators $B^{(L)}_{6}(v)$ found in \cite{Odyssey} which
generate the missing eigenstates in the algebraic Bethe ansatz of the 
six-vertex model at roots of unity.

Tools necessary to construct all eigenstates of T in the 
coordinate Bethe ansatz of the six- and 
eight-vertex models at roots of unity
have been developed by Baxter \cite{Baxter2002}. There is however a 
marked difference between our and
Baxter's result. In the coordinate Bethe ansatz \cite{Baxter2002} 
the total number of roots (the sum
of the number of regular roots and the number of roots which are 
elements of exact strings) is $\leq N$.
(see page 50 in \cite{bax733}) whereas our method 
allows the addition of an 
arbitrary number of B-strings even though the rank of the generated eigenspace
is the same in both cases. 
The six vertex model at roots of unity has a loop $sl_2$ symmetry
algebra \cite{dfm} and in  (1.37) of ref.\cite{Odyssey}   
we produced the operator $B_6^{(L)}(v)$ which 
creates $L$ strings for the six-vertex model at roots of unity  and
argued that this is the
current operator for the loop $sl_2$ symmetry algebra. This operator
has poles at the zeroes of the Laurent polynomial
\begin{equation}
Y(v)=\sum_{l=0}^{L-1}{\sinh^N{1\over 2}(v-(2l+1)i\gamma_0)\over
\prod_{k=1}^n\sinh{1\over2}(v-v_k-2il\gamma_0)
\prod_{k=1}^n\sinh{1\over2}(v-v_k-2i(l+1)\gamma_0)}
\label{drin}
\end{equation}
in $\exp(Lv)$ which we identified with the Drinfeld polynomial. 
The operator $B_6^{(L)}(v)$ is  the generator of the elements of 
the algebra in the mode basis of the loop algebra.
By differentiation of $E^{-}(z)$ (see (1.19) in \cite{Odyssey})
with respect to $z^{-1}$ and then setting $z^{-1}=0$ one can extract 
the mode operators of the 
irreducible representation related to a Bethe-state.
These conjectures
have recently been given representation theory foundation by Deguchi
\cite{deg1}.
It follows from the representation theory of the loop $sl_2$ algebra
that when the roots of the Drinfeld polynomial are distinct that the
space of states generated by $B^{(L)}_6(v)$  is a highest 
weight finite dimensional representation which is the direct sum of spin 1/2   
  representations. Accordingly the current operator $B_6^{(L)}(v)$ will be
nilpotent.

The operator 
$B^{L,1}_{l}(\lambda_c)$ in (\ref{BL1}) is, in form,  the
eight-vertex  generalization  
of the operator $B_6^{(L)}(v)$ in equ.(1.37) of
\cite{Odyssey} 
and the meromorphic function ${\hat Y}(\lambda_c)$ in
 (\ref{Y}) (which is essentially the function $G(z)$ in (5.8) of
\cite{deg3} and (31) of \cite{deg2}) appears to be  the generalization of 
the Drinfeld polynomial $Y(v)$ (\ref{drin}) of the six vertex model.
We therefore expect that the elliptic $B$-string operator 
$B^{(L,1)}_l(\lambda_c)$ is the current generator 
of the elliptic symmetry at roots of unity.  

There is, however, an important difference between   $B_6^{(L)}(v)$ and
$B^{L,1}_{l}(\lambda_c)$. We saw in the previous section that there is
no restriction on the number of $B$-strings in a state. Therefore,
in contrast to $B_6^{(L)}(v)$ the operator $B^{L,1}_{l}(\lambda_c)$ is
not nilpotent and therefore the space of states produced by the action of  
$B^{L,1}_{l}(\lambda_c)$ is not a highest weight representation as was
the case in the six vertex model. 
Here again an investigation of the six-vertex limit like in ref.\cite{DdVG} 
will give us valuable insight into the nature of symmetry.

It remains to relate the concept of the $B$-strings of this paper with
the $Q$-strings of \cite{fm4} and to compute the degeneracy of the
states from the zeroes of the meromorphic function ${\hat
  Y}(\lambda_c)$  (\ref{Y}). For the 6
vertex model it followed \cite{Odyssey} 
from the representation theory of $sl_2$ loop 
algebra that the number of degenerate states is 2 raised to the power
of the order of the Drinfeld polynomial (\ref{drin}) 
if the zeroes of the Drinfeld
polynomial are distinct. For the $Q$-strings in the eight-vertex model 
the degeneracy followed \cite{fm4} from the fact that for 
every $Q$-string center
$\lambda_c$ there was another independent $Q$-string center at 
$\lambda_c+iK'.$  These two concepts need to be united in a
representation theory of the elliptic algebra which underlies the eight-vertex
 model.

\vspace{.2in}
{\bf \large Acknowledgments}

We wish to thank Prof. R.J. Baxter,  Prof. T. Deguchi and 
Prof. G. Felder for most useful discussions.

\vspace{.2in}

\appendix

\section{Theta functions}
The definition of Jacobi Theta functions of nome $q$ is
\begin{eqnarray}
H(v)&=&2\sum_{n=1}^\infty(-1)^{n-1}q^{(n-{1\over 2})^2}\sin[(2n-1)\pi
v/(2K)]\\
\Theta(v)&=&1+2\sum_{n=1}^{\infty}(-1)^nq^{n^2}\cos(nv\pi/K)\nonumber\\
\label{thetadf}
\end{eqnarray}
where $K$  and $K'$ are the standard elliptic integrals of the first kind
and 
\begin{equation}
q=e^{-\pi K'/ K}.
\end{equation}
These theta functions satisfy the quasi periodicity 
relations (15.2.3) of ref. \cite{baxb}
\begin{eqnarray}
H(v+2K)&=&-H(v)\label{hper1}\\
H(v+2iK')&=&-q^{-1}e^{-\pi i v/K}H(v)\label{Hper}
\end{eqnarray}
and
\begin{eqnarray}
\Theta(v+2K)&=&\Theta(v)\\
\Theta(v+2iK')&=&-q^{-1}e^{-\pi i v/ K}\Theta(v).
\end{eqnarray}
$\Theta(v)$ and $H(v)$ are not
independent but satisfy
(15.2.4) of ref.\cite{baxb}
\begin{eqnarray}
\Theta (v+iK')=iq^{-1/4}e^{-{\pi i v\over 2K}}H(v)\nonumber \\
H(v+iK')=iq^{-1/4}e^{-{\pi i v\over 2K}}\Theta(v).
\label{threl}
\end{eqnarray}

\section{The algebraic Bethe ansatz}
We follow the formalism of \cite{TakFadd79} and
define what is called the monodromy matrix by
\begin{equation}
{\cal{T}} = {\cal{L}}_N \cdots {\cal{L}}_1
=
\left(\begin{array}{c c}
A & B \\
C & D \\
\end{array} \right)
\label{mono}
\end{equation}
where ${\cal{L}}_n$ is a $2 \times 2$ matrix 
in auxiliary space with entries which are
$2 \times 2$ matrices in spin space acting on 
the $n$ th spin in the spin chain and
$A,B,C,D$ are $2^{N}\times 2^{N}$ matrices in spin space.
\begin{equation}
{\cal{L}}_n = 
\left(\begin{array}{c c}
\alpha_n & \beta_n \\
\gamma_n & \delta_n\\
\end{array} \right)
\end{equation}
\begin{equation}
\alpha_n = 
\left(\begin{array}{c c}
a & 0\\
0 & b  \\
\end{array} \right),
\hspace{0.1 in}
\beta_n = 
\left(\begin{array}{c c}
0&d\\
c&0  \\
\end{array} \right)
\hspace{0.1 in}
\gamma_n = 
\left(\begin{array}{c c}
0 &c\\
d & 0  \\
\end{array} \right),
\hspace{0.1 in}
\delta_n = 
\left(\begin{array}{c c}
b & 0\\
0 & a\\
\end{array} \right)
\end{equation}
$a,b,c$ and $d$ are defined in equ.(\ref{bw8}).

A gauge transformed monodromy matrix is then defined as
\begin{equation}
{\cal{T}}_{k,l} = M^{-1}_{k}(\lambda) {\cal T(\lambda)} M_{l}(\lambda) =
\left(\begin{array}{c c}
A_{k,l} & B_{k,l} \\
C_{k,l} & D_{k,l} \\
\end{array} \right).
\label{tmono}
\end{equation}
where the  matrices $M_k$ introduced by Baxter \cite{bax732} are
\begin{equation}
M_k = 
\left(\begin{array}{c c}
x^{1}_{k} & y^{1}_{k}\\
x^{2}_{k} & y^{2}_{k}\\
\end{array} \right)
\end{equation}
\begin{equation}
\left(\begin{array}{c}
x^{1}_{k}(\lambda)\\
x^{2}_{k}(\lambda)\\
\end{array} \right)=
\left(\begin{array}{c}
H(s+2k\eta-\lambda)\\
\Theta(s+2k\eta-\lambda)\\
\end{array} \right) \hspace{0.5 in} 
\hspace{0.4 in}
\left(\begin{array}{c}
y^{1}_{k}(\lambda)\\
y^{2}_{k}(\lambda)\\
\end{array} \right)=\frac{1}{g(\tau_{k})}
\left(\begin{array}{c}
H(t+2k\eta+\lambda)\\
\Theta(t+2k\eta+\lambda)\\
\end{array} \right)
\end{equation}
\begin{equation}
 g(u) = H(u)\Theta(u), \hspace{0,3 in} \tau_l=(s+t)/2+2l\eta-K
\end{equation}
with
\begin{equation}
{\rm det}M_k={2g(\lambda+(t-s)/2)\over g(K)}\equiv m(\lambda).
\end{equation}
Thus (\ref{tmono}) is explicitly written as
\begin{equation}
A_{k,l}(\lambda)
=\frac{1}{m(\lambda)}\left(y^{2}_{k}(\lambda)x^{1}_{l}(\lambda)A(\lambda)
+y^{2}_{k}(\lambda)x^{2}_{l}(\lambda)B(\lambda)
-y^{1}_{k}(\lambda)x^{1}_{l}(\lambda)C(\lambda)-
y^{1}_{k}(\lambda)x^{2}_{l}(\lambda)D(\lambda)\right)
\label{Akl1}
\end{equation}
\begin{equation}
B_{k,l}(\lambda)
=\frac{1}{m(\lambda)}\left(y^{2}_{k}(\lambda)y^{1}_{l}(\lambda)A(\lambda)
+y^{2}_{k}(\lambda)y^{2}_{l}(\lambda)B(\lambda)
-y^{1}_{k}(\lambda)y^{1}_{l}(\lambda)C(\lambda)-
y^{1}_{k}(\lambda)y^{2}_{l}(\lambda)D(\lambda)\right)
\label{Bkl1}
\end{equation}
\begin{equation}
C_{k,l}(\lambda)
=\frac{1}{m(\lambda)}\left(-x^{2}_{k}(\lambda)x^{1}_{l}(\lambda)A(\lambda)
-x^{2}_{k}(\lambda)x^{2}_{l}(\lambda)B(\lambda)
+x^{1}_{k}(\lambda)x^{1}_{l}(\lambda)C(\lambda)+
x^{1}_{k}(\lambda)x^{2}_{l}(\lambda)D(\lambda)\right)
\label{Ckl1}
\end{equation}
\begin{equation}
D_{k,l}(\lambda)
=\frac{1}{m(\lambda)}\left(-x^{2}_{k}(\lambda)y^{1}_{l}(\lambda)A(\lambda)
-x^{2}_{k}(\lambda)y^{2}_{l}(\lambda)B(\lambda)
+x^{1}_{k}(\lambda)y^{1}_{l}(\lambda)C(\lambda)+
x^{1}_{k}(\lambda)y^{2}_{l}(\lambda)D(\lambda)\right)
\label{Dkl1}
\end{equation}
and in this notation the  transfer matrix (\ref{transfer}) becomes
\begin{equation}
T(\lambda) = A(\lambda)+D(\lambda) = A_{l,l}(\lambda)+D_{l,l}(\lambda). 
\label{transfer1}
\end{equation}
where we note that the sum $A_{l,l}(\lambda)+D_{l,l}(\lambda)$ is
independent of $l$ even though each term in the sum separately depends
on $l$.

Using relations which Baxter derived in \cite{bax732} the following 
commutation relations are obtained \cite{TakFadd79}
\begin{equation}
B_{k,l+1}(\lambda)B_{k+1,l}(\mu)=B_{k,l+1}(\mu)B_{k+1,l}(\lambda)
\label{C1}
\end{equation}
\begin{equation}
A_{k,l}(\lambda)B_{k+1,l-1}(\mu)=\alpha(\lambda,\mu)B_{k,l-2}(\mu)A_{k+1,l-1}(\lambda)
-\beta_{l-1}(\lambda,\mu)B_{k,l-2}(\lambda)A_{k+1,l-1}(\mu)
\label{C2}
\end{equation}
\begin{equation}
D_{k,l}(\lambda)B_{k+1,l-1}(\mu)=\alpha(\mu,\lambda)B_{k+2,l}(\mu)D_{k+1,l-1}(\lambda)
+\beta_{k+1}(\lambda,\mu)B_{k+2,l}(\lambda)D_{k+1,l-1}(\mu)
\label{C3}
\end{equation}
where
\begin{equation}
\alpha(\lambda,\mu)=\frac{h(\lambda-\mu-2\eta)}{h(\lambda-\mu)},
~~~{\rm and}~~~
\beta_k(\lambda,\mu)=\frac{h(2\eta)h(\tau_k+\mu-\lambda)}{h(\tau_k)h(\mu-\lambda)}
\label{alpha}
\end{equation}
and where
\begin{equation}
h(u)=\Theta(0)\Theta(u)H(u)
\end{equation}
The commutation relations (\ref{C1})-(\ref{C3}) are valid for 
all values of $\eta$.

In order to proceed further a set of direct product vectors is defined by
\begin{equation}
\Omega_{N}^{l}=\omega_1^{l}\otimes \cdots \otimes \omega_N^{l}
\label{vac}
\end{equation}
\begin{equation}
\omega_n^{l}=
\left(\begin{array}{c c}
H(s+2(n+l)\eta-\eta)\\
\Theta(s+2(n+l)\eta-\eta)\\
\end{array} \right)
\label{localvac}
\end{equation}
and it is shown \cite{TakFadd79} that
\begin{equation}
C_{N+l,l}\Omega_{N}^{l} = 0
\end{equation}
\begin{equation}
A_{N+l,l}\Omega_{N}^{l} = h^{N}(\lambda+\eta)\Omega_{N}^{l-1}
\end{equation}
\begin{equation}
D_{N+l,l}\Omega_{N}^{l} = h^{N}(\lambda-\eta)\Omega_{N}^{l+1}
\end{equation}
Finally define the set of vectors
\begin{equation}
\psi_{l}(\lambda_1,\cdots,\lambda_n)
=B_{l+1,l-1}(\lambda_1)\cdots B_{l+n,l-n}(\lambda_n)
\Omega_{N}^{l-n}
\end{equation}
Then for the case that $\eta$ satisfies the root of unity condition
(\ref{root}) and $N$ is even  
one obtains the eigenstates
\begin{equation}
\Psi_{m}
=\sum_{l=0}^{L-1}e^{2\pi iml/L}
\psi_{l}(\lambda_1,\cdots,\lambda_n)
\label{help100}
\end{equation} 
where $\lambda_1, \cdots , \lambda_n$ are chosen to satisfy what are
called Bethe's equations 
\begin{equation}
\frac{h^{N}(\lambda_j+\eta)}{h^{N}(\lambda_j-\eta)}=
e^{-4\pi im/L}\prod_{k=1,k\neq j}^{n}
\frac{h(\lambda_j-\lambda_k+2\eta)}{h(\lambda_j-\lambda_k-2\eta)}.
\label{Betheq}
\end{equation}
with $N=2n+$integer$\times L$ 
and $m=0,1,\cdots ,L-1$.\\
It is important to note that all equations are valid for generic
values of $\eta/K$ except for the final expression (\ref{help100}) for the
eigenvectors which is the only place where the root of unity condition
(\ref{root}) is used.

\end{document}